\documentclass[twocolumn,aps,prb,superscriptaddress]{revtex4}
\usepackage{epsfig,amsmath,amssymb,wasysym}
\usepackage{hyperref}

\newcommand{\sbar}{\bar{s}}

\newcommand{\be}{\begin{equation}}
\newcommand{\ee}{\end{equation}}
\newcommand{\bk}{{{\bf{k}}}}
\newcommand{\bq}{{{\bf{q}}}}
\newcommand{\bp}{{{\bf{p}}}}

\newcommand{{\bQ}}{{{\bf{Q}}}}
\newcommand{\br}{{{\bf{r}}}}

\newcommand{\bea}{\begin{eqnarray}}
\newcommand{\eea}{\end{eqnarray}}
\newcommand{\ra}{\rangle}
\newcommand{\la}{\langle}

\newcommand{\bS}{{\bf S}}

\newcommand{\dg}{{\dagger}}
\newcommand{\pdg}{{\phantom\dagger}}
\newcommand{\nn}{\nonumber}

\newcommand{\bdelta}{{{\bf{\delta}}}}

\begin{document}
\title{N\'eel to dimer transition in spin-$S$ antiferromagnets: Comparing 
bond operator theory with quantum Monte Carlo simulations for bilayer 
Heisenberg models}
\author{R. Ganesh}
\affiliation{Department of Physics, University of Toronto, Toronto, Ontario, Canada M5S 1A7}
\author{Sergei V. Isakov}
\affiliation{Theoretische Physik, ETH Zurich, 8093 Zurich, Switzerland}
\author{Arun Paramekanti}
\affiliation{Department of Physics, University of Toronto, Toronto, Ontario, Canada M5S 1A7}
\affiliation{Canadian Institute for Advanced Research, Toronto, Ontario, Canada M5G 1Z8}
\date{\today}

\begin{abstract}
{We study the N\'eel to dimer transition driven by interlayer exchange coupling in spin-$S$ Heisenberg 
antiferromagnets on bilayer square and honeycomb lattices for $S\!\!=\!\!1/2$, $1$, $3/2$.
Using exact stochastic series expansion quantum Monte Carlo (QMC) calculations, we find that the 
critical value of the interlayer coupling, $J_{\perp c}[S]$, increases with
increasing $S$, with clear evidence that the transition is in the $O(3)$
universality class for all $S$. Using bond operator mean field theory restricted 
to singlet and triplet states, we find $J_{\perp c} [S] \!\propto\! S(S+1)$, in qualitative accord with 
QMC, but the resulting $J_{\perp c} [S]$ is significantly smaller than the QMC
value. For $S\!\!=\!\!1/2$, incorporating triplet-triplet interactions within a variational approach yields
a critical interlayer coupling which agrees well with QMC. For
higher spin, we argue that it is crucial to account for the high energy
quintet modes, and show that including these within a perturbative
scheme leads to reasonable
agreement with QMC results for $S\!\!=\!\!1$,$3/2$. We discuss the broad implications of our results for systems
such as the triangular lattice $S\!\!=\!\!1$ dimer compound Ba$_3$Mn$_2$O$_8$ and the $S\!=\!3/2$
bilayer honeycomb material Bi$_3$Mn$_4$O$_{12}$(NO$_{3}$).}
\end{abstract}
\maketitle

\section{Introduction}
Spin dimer compounds provide the simplest realization of a magnetically disordered 
ground state --- one where strongly coupled pairs of spins entangle to form singlets. 
Such systems are also of 
great interest since they undergo magnetic field induced spin ordering via a quantum phase 
transition which is analogous to Bose-Einstein condensation.\cite{Nikuni2000,ricereview,tchernyshyovreview,sachdevbook} There are many well-known 
spin dimer compounds\cite{Oosawa1999,hanpurple2004,organicbilayer2006}
and well studied model Hamiltonians\cite{Millis1993,sandvik1994,spinwave1995,TylerDodds} exhibiting such  
physics for $S\!\!=\!\!1/2$ spins. However, ongoing experiments on higher spin systems, such as the $S\!\!=\!\!1$ 
triangular
lattice dimer compound\cite{Stone2008} 
Ba$_3$Mn$_2$O$_8$, point to a need
to better understand higher spin generalizations and instabilities of such dimer states
driven by inter-dimer interactions.

Here, we explore this issue using a simple model which exhibits such a 
dimerized ground state - the bilayer Heisenberg antiferromagnet with a Hamiltonian given by
\bea
H = J_\perp \sum_{i}\bS_{i,1}\cdot\bS_{i,2} + J_1 \sum_{\la ij\ra}\sum_{\ell=1,2} \bS_{i,\ell}\cdot \bS_{j,\ell} .
\label{genHamiltonian}
\eea
Here,  $i$ labels sites in one layer, $\ell=1,2$ is the layer index, and $\la ij\ra$ represents nearest
neighbor pairs of spins within each layer. For $J_\perp \gg J_1$, 
the first term in $H$ dominates and
the ground state is composed of isolated interlayer singlets with $\bS_{i,1} + \bS_{i,2}=0$  for every $i$. 
If $J_\perp \ll J_1$, the system will order magnetically provided the second (intralayer)
term in the Hamiltonian is not too frustrated by the lattice geometry.
Here, we restrict our attention to cases where each layer is itself 
a bipartite lattice so that the ground
state for $J_\perp \ll J_1$ has long-range N\'eel order.
This model Hamiltonian has been extensively studied for the $S=1/2$ square lattice bilayer\cite{sandvik1994,matsushita1999,Yu1999,Capponi2007,Tripletwave2008,RVB2011}.
Effects of disorder, induced by site dilution, have also been explored.\cite{disorderedbilayer2005}
However, there has been relatively little work on understanding the higher spin generalizations of the Hamiltonian in Eq. \ref{genHamiltonian}. 
The spin-S square lattice bilayer has been studied using Schwinger boson mean field theory\cite{Ng1996} and series expansions.\cite{Gelfand1998} Variants of this spin-S model have been argued to support novel spin solid phases in three dimensions.\cite{Motrunich2007}
In this paper, 
we study this Hamiltonian for $S=1/2,1,3/2$ spins on square and honeycomb bilayers using
exact quantum Monte Carlo simulation algorithms\cite{sse} and approximate analyses based on a bond operator 
method generalized
to arbitrary spin.\cite{brijesh2010}

\begin{figure}[tb]
\includegraphics[width=\columnwidth]{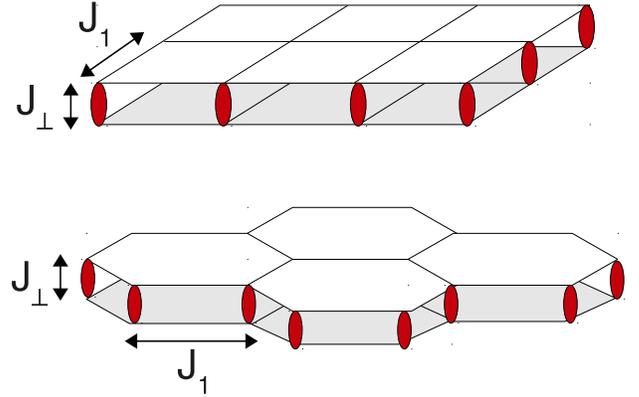}
\caption{The interlayer dimer state on square and honeycomb bilayers with singlet correlations between layers. The in-plane exchange $J_1$ and the interplane exchange $J_\perp$ act as shown.}
\label{fig.ilvbscartoon}
\end{figure}

Our main results are as follows. 
(i) Using exact stochastic series expansion quantum Monte Carlo (QMC) calculations, we 
find a N\'eel to dimer transition with increasing $J_\perp$ that is in the $O(3)$ universality class
for all the models we have studied. (ii) The critical value of the interlayer coupling, $J_{\perp c} [S]$, 
for the N\'eel to dimer transition is found to increase for higher spin. Using a bond operator mean field 
theory restricted  to singlet and triplet states, we find
$J_{\perp c} [S] \propto S(S+1)$, in qualitative accord with 
QMC results. However, there is a quantitative discrepancy between the mean field 
$J_{\perp c} [S]$ and its QMC value, which becomes more significant for higher spin.
(iii) For $S=1/2$, we show that taking into account triplet-triplet interactions within a variational approach
brings the $J_{\perp c} [S]$ value close to the QMC result. For higher spin, we show 
that the dominant corrections to the critical point arise from the high energy quintet modes
and direct triplet-triplet interactions are less important.
Incorporating the quintet excitations within a perturbative 
treatment is shown to yield a critical interlayer coupling which is in good agreement with 
QMC results for $S=1$,$3/2$. We discuss the broad implications of our results for high spin antiferromagnets such 
as the triangular lattice $S=1$ dimer compound\cite{Stone2008} Ba$_3$Mn$_2$O$_8$ and the $S=3/2$
bilayer honeycomb material\cite{BMNO} Bi$_3$Mn$_4$O$_{12}$(NO$_{3}$).

This paper is organized as follows. Section II contains results from the QMC simulations on the phase
diagram of the honeycomb and square lattice bilayer models for $S=1/2$, $1$, $3/2$. In Section III, 
we outline the bond operator formalism generalized to the case of spin-S. Section IV gives bond operator mean field theory results for the square and honeycomb lattice models. Section V discusses the variational approach that we use to take into account corrections beyond mean field theory. Section VI analyses the $S=1/2$
model including the effect of triplet-triplet interactions, while Section VII
contains a treatment of the dominant quintet corrections for $S > 1/2$. We end with a discussion in Section VIII. Details are contained in Appendices.

\section{Quantum Monte Carlo simulations}

\begin{figure}[ht]
\includegraphics[width=\columnwidth]{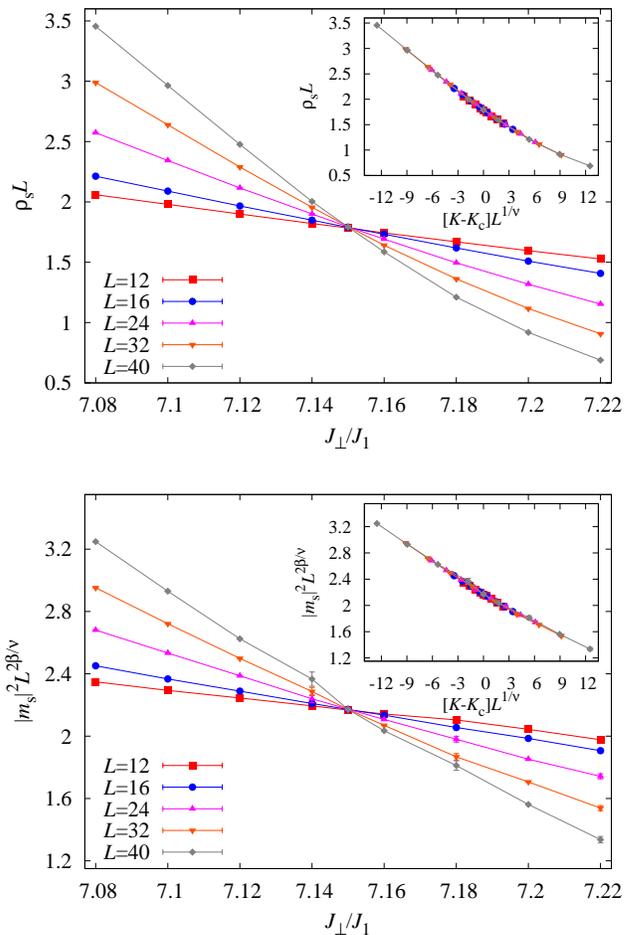}
\caption{ Scaling of the superfluid density (upper panel) and of the staggered
magnetization density squared (lower panel) for the $S=1$ antiferromagnet on
the bilayer square lattice. The curves cross at a distinct point around
$J_\perp/J_1=7.15$. The insets show the corresponding data collapse for
$z=1$, $\nu=0.7112$, $\beta=0.3689$, and $J_{\perp c}/J_1=7.15$. Lines guide
the eye. The error bars are smaller than the symbol size if not visible.}
\label{fig:squ:scaling:2}
\end{figure}

\begin{figure}[ht]
\includegraphics[width=\columnwidth]{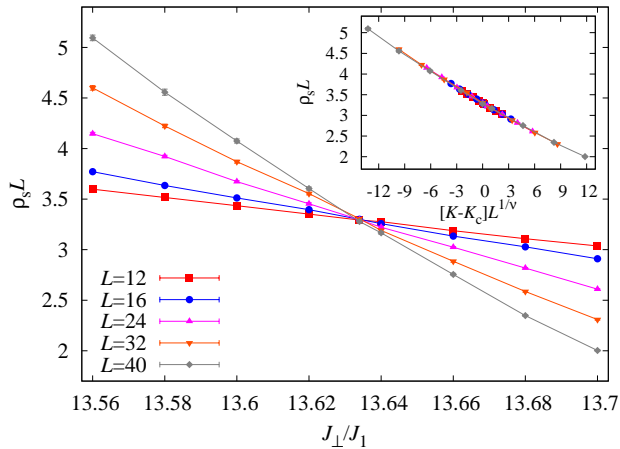}
\caption{ Scaling of the superfluid density for the $S=3/2$ antiferromagnet on
the bilayer square lattice. The curves cross at a distinct point around
$J_\perp/J_1=13.634$. The inset shows the corresponding data collapse for
$z=1$, $\nu=0.7112$, and $J_{\perp c}/J_1=13.634$.
Lines guide the eye. The error bars are smaller than the symbol size if not
visible.}
\label{fig:squ:scaling:3}
\end{figure}

The bilayer honeycomb and square lattices are bipartite lattices which can be split into two sublattices A
and B with every lattice bond being a link between sites belonging to different
sublattices.
This ensures that there is no sign problem, so that the model in Eq.~\ref{genHamiltonian}
is amenable to quantum Monte
Carlo simulations.
We perform quantum Monte Carlo simulations for $S=1/2,1,3/2$ on the bilayer
square (of linear system size $L=12,16,24,32,40$) and bilayer honeycomb
($L=12,18,24,30,36$) lattices using the Stochastic Series Expansion
algorithm.\cite{sse} For $S=1$ and $S=3/2$, simulations are performed with
modified worm weights, which lead to a slightly more efficient algorithm,
 as in Ref.~\onlinecite{sse2}.
At large enough ratio $J_\perp/J_1$, the system undergoes a quantum phase
transition from a N\'eel state to a dimerized paramagnetic state. To locate
quantum critical points, we perform finite size scaling analysis of
the superfluid density and the staggered magnetization density
squared. We measure the superfluid density $\rho_s$ by measuring winding
number fluctuations\cite{rhos}
$$
 \rho_s=T\frac{\langle W_1^2+W_2^2\rangle}{2 J_1},
$$
where $W_{1,2}$ are the winding numbers in two spatial directions and $T$
is the temperature. The staggered magnetization density squared is
given by
$$
 |m_s|^2=3 \left[\frac{1}{N}\sum_i^N (-1)^p S^z_i \right]^2,
$$
where $(-1)^p=1$ for lattice sites from sublattice A, $(-1)^p=-1$ for sites
from sublattice B, and $N$ is the number of lattice sites. In the vicinity
of a continuous phase transition the superfluid density scales as
$$
 \rho_s=L^{2-d-z}F([K-K_c]L^{1/\nu},\frac{1}{TL^z}),
$$
where $L$ is the linear system size, $d=2$ is the dimensionality of the system,
$T$ is the temperature, $[K-K_c]\equiv [(J_\perp-J_{\perp c})/J_1]$ is
the distance from the critical point $J_{\perp c}/J_1$, and $\nu$ is
the correlation length critical exponent. The staggered magnetization density
squared scales as
$$
 |m_s|^2=L^{-2\beta/\nu} M([K-K_c]L^{1/\nu},\frac{1}{TL^z}),
$$
where $\beta$ is the critical exponent. If one plots $\rho_s L^z$ as
a function of $J_\perp/J_1$ at large enough and fixed value of $1/(TL^z)$
then the curves for different system sizes should cross at the critical
point $J_{\perp c}/J_1$. If one plots $\rho_s L^z$ as a function of
$[K-K_c]L^{1/\nu}$, with appropriately chosen values of the critical
exponents and $K_c$, the curves for different systems sizes should collapse
onto the universal curve given by the function $F$. Similarly,
$|m_s|^2 L^{2\beta/\nu}$ as a function of $J_\perp/J_1$ should have
a distinct crossing point at the critical point and $|m_s|^2 L^{2\beta/\nu}$
as a function of  $[K-K_c]L^{1/\nu}$ should collapse onto the universal curve
given by the function $M$. We perform simulations at fixed aspect ratio
$T=J_1/2L$.

\begin{figure}[ht]
\includegraphics[width=\columnwidth]{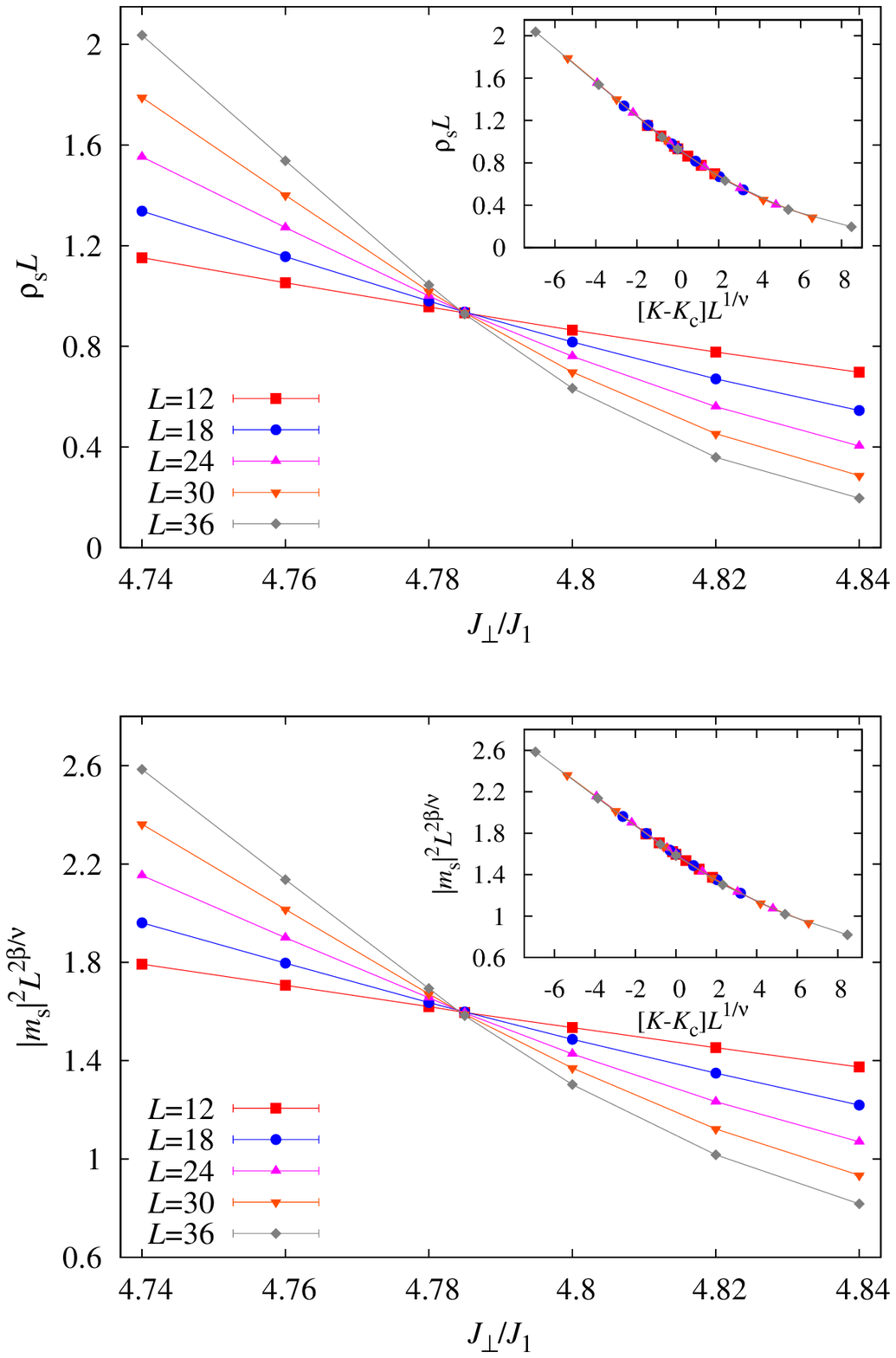}
\caption{ Scaling of the superfluid density (upper panel) and of the staggered
magnetization density squared (lower panel) for the $S=1$ antiferromagnet on
the bilayer honeycomb lattice. The curves cross at a distinct point around
$J_\perp/J_1=4.785$. The insets show the corresponding data collapse for $z=1$,
$\nu=0.7112$, $\beta=0.3689$, and $J_{\perp c}/J_1=4.785$. Lines guide the eye.
The error bars are smaller than the symbol size if not visible.}
\label{fig:hex:scaling:2}
\end{figure}

\begin{figure}[ht]
\includegraphics[width=\columnwidth]{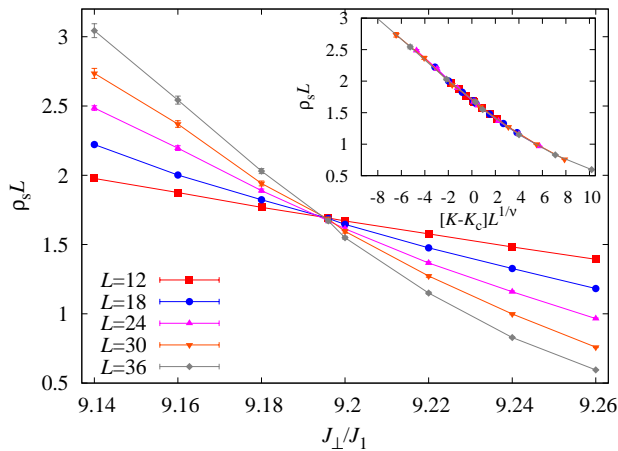}
\caption{ Scaling of the superfluid density for the $S=3/2$ antiferromagnet on
the bilayer honeycomb lattice. The curves cross at a distinct point around
$J_\perp/J_1=9.194$. The inset shows the corresponding data collapse for
$z=1$, $\nu=0.7112$, and $J_{\perp c}/J_1=9.194$. Lines guide
the eye. The error bars are smaller than the symbol size if not visible.}
\label{fig:hex:scaling:3}
\end{figure}

\subsection{Square lattice}

The quantum critical point for the $S=1/2$ bilayer quantum antiferromagnet on
the square lattice was found in Refs.~\onlinecite{sandvik1994,squbilayer}
$J_{\perp c}/J_1=2.5220(1)$. In the present work, we find that the quantum
critical points are located at $J_{\perp c}/J_1=7.150(2)$ for $S=1$, and
at $J_{\perp c}/J_1=13.634(3)$ for $S=3/2$. The data scale
very well with the critical exponents $\nu=0.7112$ and $\beta=0.3689$ of
the $O(3)$ universality class\cite{heisexp} for any value of spin.
The crossing points and data collapse for $S=1$ and $S=3/2$ are shown
in Figs.~\ref{fig:squ:scaling:2} and~\ref{fig:squ:scaling:3}.
Note that we do not show the scaling of the magnetization density squared for
$S=3/2$ because the data points are too noisy.

\subsection{Honeycomb lattice}

We find that that for the honeycomb lattice the quantum critical points are
located at $J_{\perp c}/J_1=1.645(1)$ for $S=1/2$, $J_{\perp c}/J_1=4.785(1)$
for $S=1$, and
$J_{\perp c}/J_1=9.194(3)$ for $S=3/2$.
The data scale very well with the critical exponents  $\nu=0.7112$ and
$\beta=0.3689$ of the $O(3)$ universality class\cite{heisexp} for any value
of spin. The crossing points and data collapse for $S=1$ and $S=3/2$ are shown
in Figs.~\ref{fig:hex:scaling:2} and~\ref{fig:hex:scaling:3}.
We do not show the scaling of the magnetization density squared for
$S=3/2$ because the data points are too noisy.

\begin{figure}[ht]
\includegraphics[width=\columnwidth]{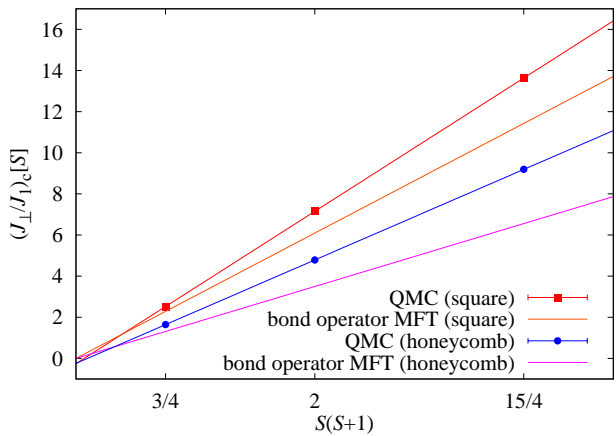}
\caption{ $J_{\perp c}/J_1[S]$ as a function of $S(S+1)$ for the bilayer
square and honeycomb lattices. Lines are linear fits. Note that the curves
cross approximately at $S(S+1)=0$.}
\label{fig:qcp}
\end{figure}

\subsection{Critical point as a function of spin}

In Fig.~\ref{fig:qcp}, we show the quantum critical points $J_{\perp c}/J_1$
as functions of $S(S+1)$ for the bilayer square and honeycomb lattices.
We find that $J_{\perp c}/J_1[S]$ is a linear function of $S(S+1)$. In the following
sections, we will attempt to make sense of these QMC results using an extension
of the bond operator theory of the dimerized state and its instability to N\'eel order.

\section{Bond operator representation}
An elegant
approach which allows us to understand the physics of the dimer ground state and its magnetic
ordering instabilities is the bond operator formalism which was first proposed for $S=1/2$ 
antiferromagnets. \cite{sachdevbhatt1990}  In this scheme, the 
spin operators are
represented in a new basis consisting of singlet and triplet states on the interlayer bonds $(i,1)$-$(i,2)$.  
In the limit where the intralayer coupling $J_1=0$, the ground state
consists of localized singlets on these bonds, with a gap $J_\perp$ to the triplet
excitations. A nonzero $J_1$ allows a pair of neighboring bonds $(i,1)$-$(i,2)$ and $(j,1)$-$(j,2)$ to
exchange their singlet/triplet character. Such a `triplet hopping' process converts the localized triplet
modes into dispersing `triplons', with three-fold degenerate bands due to the underlying $SU(2)$
symmetry of the Hamiltonian.
In this picture, the dimer to N\'eel transition is an $O(3)$ transition driven by the condensation of triplon
modes at a certain wavevector where the dispersion minimum hits zero. Generalizations of this approach to spin-1 magnets have been proposed earlier.\cite{Chubukov1991,Wang2000} 
Here, we adopt a recent generalization of the bond operator method to arbitrary spin \cite{brijesh2010} to study 
bilayer Heisenberg antiferromagnets.

In a spin-$S$ bilayer system, in the limit $J_\perp \gg J_1$, we have isolated interlayer bonds. The bond can be in one of the following states: a singlet, a $3$-fold degenerate triplet, a $5$-fold
quintet, etc. We introduce one boson for each of these states:
\bea
\nn \vert s_i \ra &\equiv& s_i^\dg \vert 0 \ra, \\
\nn \vert t_{i,m\in\{-1,0,1\}} \ra &\equiv& t_{i,m}^\dg \vert 0 \ra,\\
\nn \vert q_{i,m\in\{-2,\cdots,2\}} \ra &\equiv& q_{i,m}^\dg \vert 0 \ra, \\
\nn &\vdots&
\eea
The index i here runs over all interlayer bonds, and $m$ 
labels the $S_z$-component of the total spin on the
interlayer bond. These boson operators form the basis for a bond operator
representation. 
To restrict to the physical Hilbert space of spins, every interlayer bond should have exactly one boson,
\bea
s_i^\dg s_i + \sum_{m=-1,0,1} t_{i,m}^\dg t_{i,m} + \sum_{n=-2,\cdots,2} q_{i,n}^\dg q_{i,n} +\cdots= 1.
\label{eq:constraint}
\eea
In terms of bond operators, the exchange interaction on an interlayer bond is given by
\bea
\label{ILexch}
J_\perp~\bS_{i,1} \cdot \bS_{i,2} &=& \varepsilon_s s_i^\dg s_i
+ \varepsilon_t \sum_{m=-1,0,1} t_{i,m}^\dg t_{i,m} \nn \\
&+&\varepsilon_q \sum_{m=-2,\cdots,2} q_{i,m}^\dg q_{i,m}+\cdots
\eea
where  $\varepsilon_s = -J_\perp S(S+1)$, 
$\varepsilon_t = J_\perp \{ 1-S(S+1) \}$, and $\varepsilon_q = J_\perp\{3-S(S+1)\}$.

Bond operator theory re-expresses the spin operators and their interactions in terms of these bond bosons.
In the limit $J_\perp \gg J_1$, the singlets, triplets, quintets, etc. form a hierarchy with the energy spacing between 
each tier of order $J_\perp$. In this paper, we restrict our analysis to the low energy subspace of singlets, triplets 
and quintets on a bond, and neglect higher spin states as they are much higher in energy.

We first turn to the usual bond operator mean field theory retaining only singlet and triplet modes,
ignoring triplet interactions and higher excited states and imposing the 
constraint in Eq.~\ref{eq:constraint} on average.
We then consider, in turn, the effect of triplet-triplet interactions for $S=1/2$ and the effect of quintet states for
$S > 1/2$. For convenience of notation, we henceforth set $J_1=1$,
thus measuring $J_\perp$ in units of $J_1$.

\section{Singlet-Triplet mean field theory}
At mean field level, the interlayer dimer state is described by a uniform condensate of the singlet bosons, with 
$\la s_i \ra= \la s_i^\dg \ra=\sbar$. Retaining only triplet excitations, the spin operators at each site are given by \cite{brijesh2010}
\bea
\nn S^{+}_{i,\ell} &=& (-1)^\ell\sqrt{\frac{2S(S+1)}{3}}
\sbar\{ t_{i,-1} -t_{i,1}^\dg \} \\
&+& \frac{1}{\sqrt{2}}\{ t_{i,1}^\dg t_{i,0}+t_{i,0}^\dg t_{i,-1} \}, \\
\nn S_{i,\ell}^z &=& (-1)^\ell \sqrt{\frac{S(S+1)}{3}}
\sbar \{ t_{i,0}+t_{i,0}^\dg \} \\
&+&\frac{1}{2} \{ t_{i,1}^\dg t_{i,1} - t_{i,-1}^\dg t_{i,-1}\}.
 \eea
 
Using these expressions, the Hamiltonian takes the form
\bea
 H _{\rm mf} \!\!&\!\!=\!\!&\!\!  \varepsilon_s N_\perp \sbar^2  \!+\! \varepsilon_t \! \sum_{i, m} \! 
 t_{i,m}^\dg t^\pdg_{i,m} \! \! -\! \mu \! \sum_i \! \left( \sum_m t_{i,m}^\dg t^\pdg_{i,m} \! +\! \sbar^2 \!- \!1\right) \nn \\
\nn &+& \!\!\! \frac{2S(S+1)}{3}\sbar^2\sum_{\la i, j \ra}\Big[ \{ t_{i,0}+t_{i,0}^\dg \}\{ t_{j,0}+t_{j,0}^\dg \} \\
&+& \Big( \{ t_{i,-1}\!-\!t_{i,1}^\dg \}\{t_{j,-1}^\dg \!-\! t_{j,1}\}
+h.c.\Big)\Big],
\label{Hmft}
\eea
where $\mu$ is a Lagrange
multiplier which enforces the constraint in Eq.~\ref{eq:constraint} on average. $N_\perp$ is the number
of interlayer bonds. We have dropped
quartic terms in the triplet operators (which corresponds to ignoring
triplet-triplet interactions).

In the rest of this paper, we use the following two basis sets to represent triplet states: $\{ \vert t_{-1} \ra_i, \vert t_{0} \ra_i, \vert t_{1} \ra_i\}$ or $\{ \vert t_{x} \ra_i, \vert t_{y} \ra_i, \vert t_{z} \ra_i\}$. The former basis labels states by the z-projection of spin. The latter labels each state by the direction in which its spin projection is zero. We can go from one basis to another using $\vert t_{0} \ra_i = \vert t_{z} \ra_i$ and $\vert t_{\pm 1} \ra_i = (\mp\vert t_x \ra_i - i\vert t_y \ra_i)/\sqrt{2} $. Below, we 
will use the index $m$ to represent an element of the first basis and $u$ to represent an element of the second.

\begin{figure}[tb]
\includegraphics[width=2in]{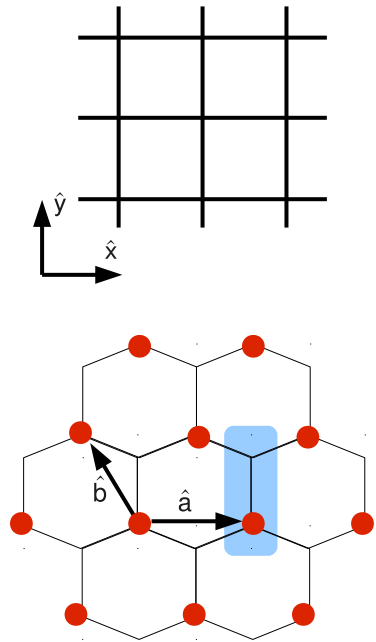}
\caption{Top view of bilayers. (Top) Square lattice with primitive lattice vectors $\hat{x}$ and $\hat{y}$ shown.
(Bottom) Honeycomb lattice. The shaded region is the unit cell composed of two sites. Sites marked with a red circle belong to the A sublattice. Unmarked sites belong to the B sublattice. The primitive lattice vectors $\hat{a}$ and $\hat{b}$ are shown. 
}
\label{fig:lattices}
\end{figure}

\subsection{Square lattice bilayer}

%\begin{figure}[tb]
%\includegraphics[width=\columnwidth]{sqilvbswlabels.eps}
%\caption{Square lattice bilayer system. In-plane exchange J$_1$ and interlayer coupling J$_\perp$ act as indicated. The primitive vectors %$\hat{x}$ and $\hat{y}$ are shown. }
%\label{fig:squarelatt}
%\end{figure}
%
A top view of the square lattice bilayer with the relevant primitive lattice vectors
is shown in Fig.~\ref{fig:lattices}.
At mean field level, the Hamiltonian of Eq.~\ref{genHamiltonian} may be written as 
\bea
\label{MFTHamiltonian}
\nn H^{(0)}_{_{\Box}} = -J_\perp N_\perp S(S+1)\sbar^2 -\mu \sbar^2 N_\perp + \mu N_\perp -\frac{3N_\perp A}{2}\\
+\sum_{\bk,u\in\{x,y,z\}}\!\!\!\!\!\!\!\!{}^{'}\phantom{ab}
\psi_{\bk,u}^\dg
\left(\begin{array}{cc}
A+2\epsilon_\bk & 2\epsilon_\bk \\
2\epsilon_\bk & A+2\epsilon_\bk \end{array}\right)
\psi_{\bk,u}, 
\eea
where $\psi_{\bk,u}=[t_{\bk,u}\phantom{ab} t_{-\bk,u}^\dg]^T$. The primed summation indicates that if $\bk$ is included in the sum, then $-\bk$ is excluded. The coefficients in the Hamiltonian matrix are
\bea
A &=& J_\perp \{1-S(S+1)\} -\mu,\\
\epsilon_\bk &=& \frac{2S(S+1)}{3}\sbar^2(\cos(k_x)+\cos(k_y)).
\eea
Diagonalizing this Hamiltonian matrix by a bosonic Bogoliubov transformation (see Appendix \ref{app.sqBog}), we obtain eigenvalues 
$\lambda_\bk=\sqrt{A(A+4\epsilon_\bk)}$ for the energies of the independent `triplon' modes. Each of these modes 
adds a zero point contribution to the ground state energy, yielding
\bea
\nn E_{_{\Box}}^{(0)} &=& -J_\perp N_\perp S(S+1)\sbar^2 -\mu \sbar^2 N_\perp + \mu N_\perp \\
&-& \frac{3N_\perp A}{2}
+3\sum_{\bk}{}^{'}\lambda_\bk.
\label{E2GS}
\eea
We minimize this ground state energy with respect to $\mu$ and $\sbar$, via
$\partial E_{_\Box}^{(0)}/\partial \mu=0$ and $\partial E_{_\Box}^{(0)}/\partial \sbar^2=0$.
This yields the two equations
\bea
\label{mftconstraint}
\sbar^2 &=& \frac{5}{2} - \frac{3}{N_\perp}\sum_{\bk}{}^{'}\frac{A+2\epsilon_\bk}{\lambda_\bk}, \\
\label{egymin}
\mu &=&  -J_\perp S(S+1)+\frac{6}{N_\perp}\sum_{\bk}{}^{'}\frac{A\epsilon_\bk}{\sbar^2 \lambda_\bk}.
\eea
Using the values of $\sbar$ and $\mu$ thus obtained, we may calculate the gap to triplet excitations.
The dimer-N\'eel transition occurs when the triplon gap vanishes at $J_\perp=J_{\perp c}$. 
We have explicitly checked that triplon
condensation at $\bk=(\pi,\pi)$  yields N\'eel order on the bilayer.
Using 
Eqns.~\ref{mftconstraint},\ref{egymin} above, we arrive at the following two results at the critical point.
(i) The value $\sbar$ at the dimer-N\'eel critical point is independent of spin and is given by 
\bea
\sbar_{\rm c}^2 = \frac{5}{2}-\frac{3}{2N_\perp}\sum_\bk {}^{'}
\frac{4+(\cos k_x +\cos k_y)}{\sqrt{4+2(\cos k_x + \cos k_y)}}.
\eea
A numerical evaluation shows $\sbar_c \approx 0.904$.
(ii) We find the location of the dimer-N\'eel critical point
\bea
\label{jccrit}
\! J_{\perp c}\!\! =\!\! S(S\!\!+\!\! 1)\!\!\left[\! \frac{40}{3}\!-\!\frac{32}{N_\perp}\! \sum_{\bk}{}^{'}\!\!\! \frac{1}{\sqrt{4+2(\cos k_x +\cos k_y )}}\! \right]\!\!.
\eea
A numerical evaluation yields $J_{\perp c} \approx 3.047 S(S+1)$.
For $S=1/2$, this mean field result, $J_{\perp c}[S=1/2] \approx 2.286$, agrees with previous work\cite{matsushita1999} and is slightly smaller than the
QMC value. \cite{sandvik1994} For higher spin, the mean field estimates,
$J_{\perp c}[S=1] \approx 6.095$ and $J_{\perp c}[S=3/2] \approx 11.428$, are significantly smaller than our
QMC results. This comparison is summarized in Table \ref{Table:square}.
The scaling result $J_{\perp c} \sim S(S+1)$ has been suggested in Ref. \onlinecite{Gelfand1998} on the basis of series expansion calculations.
Remarkably, as shown in Fig.\ref{fig:qcp},
this scaling relation derived from mean-field theory seems to be reasonably accurate
even for exact QMC results.

\subsection{Honeycomb lattice bilayer}

%\begin{figure}[tb]
%\includegraphics[width=\columnwidth]{hcilvbswlabels.eps}
%\caption{Top: Single honeycomb layer. The shaded region is the unit cell composed of two sites. Sites marked with a red circle belong to the A sublattice. Unmarked sites belong to the B sublattice. The primitive lattice vectors $\hat{a}$ and $\hat{b}$ are shown. 
%Bottom: Cartoon picture of the interlayer dimer state. In-plane exchange $J_1$ (set to unity in the paper) and interlayer coupling $J_\perp$ are shown. Interlayer bonds possess singlet correlations as shown.}
%\label{fig:honeycomblatt}
%\end{figure}

The honeycomb lattice is composed of two interpenetrating triangular lattices, as shown in Fig.\ref{fig:lattices}. Operators therefore come with an additional sublattice index which distinguishes A and B sublattices.
The mean field Hamiltonian is given by
\bea
\label{hcmfthamiltonian}
\nn H^{(0)}_{\hexagon} &=& -N_\perp J_\perp S(S+1)\sbar^2 -N_\perp\mu \sbar^2  + N_\perp\mu \\
&-& \frac{3N_\perp C}{2}
+\sum_{\bk,u}{}^{'}
\psi_{\bk,u}^\dg M_\bk \psi_{\bk,u}, \phantom{ab}
\eea
where $C=(J_\perp\{1-S(S+1)\}-\mu)$. $N_\perp$ denotes the number of interlayer bonds in the honeycomb bilayer. 
The operator $\psi_{\bk,u}$ and the Hamiltonian matrix $M_\bk$ are given by
\bea
\label{Eq.hcmftpsiM}
\psi_{\bk,u}=\left(\begin{array}{c}
t_{\bk,A,u} \\ t_{\bk,B,u} \\ t_{-\bk,A,u}^\dg \\  t_{-\bk,B,u}^\dg 
\end{array}\right), \phantom{ab}
M_\bk =
\left(\begin{array}{cccc}
C & \beta_\bk & 0 & \beta_\bk \\
\beta_\bk^* & C & \beta_\bk^* & 0 \\
0 & \beta_\bk & C & \beta_\bk \\
\beta_\bk^* & 0 & \beta_\bk^* & C
\end{array}\right),
\eea
where $\beta_\bk=2 \frac{S(S+1)}{3}\sbar^2 \gamma_\bk$, with $\gamma_\bk= 1+e^{-ik_b}+e^{-ik_a-ik_b}$, and
we have defined $k_a \equiv \bk\cdot \hat{a}$ and $k_b \equiv \bk\cdot \hat{b}$.
Diagonalizing this Hamiltonian (see Appendix \ref{app.hcBog}), we obtain two eigenvalues for every $\bk$. The eigenvalues are given 
by $\lambda_{\bk,1/2}=\sqrt{C^2 \mp 2C\vert \beta_\bk \vert}$.
The mean field ground state energy is given by 
\bea
\nn E_{\hexagon}^{(0)} &=& -N_\perp J_\perp S(S+1)\sbar^2 -N_\perp \mu \sbar^2  + N_\perp \mu \\
&-& \frac{3N_\perp C}{2}
+3\sum_{\bk}{}^{'} (\lambda_{\bk,1}+\lambda_{\bk,2}).
\label{hcE2GS}
\eea
As before, we demand
$\partial E^{(0)}_{\hexagon}/\partial \mu=0$ and $\partial E^{(0)}_{\hexagon}/\partial \sbar^2=0$.
This leads to the two mean field equations
\bea
\label{Eq.hcMFTmu}
\sbar^2 &=& \frac{5}{2} - \frac{3}{N_\perp}
\sum_{\bk}{}^{'} \left[\frac{C-|\beta_\bk|}{\lambda_{\bk,1}}
+ \frac{C + |\beta_\bk|}{\lambda_{\bk,2}}\right], \\
\nn \mu &=& - \frac{2CS(S+1)}{N_\perp}
\sum_{\bk}\!{}^{'} \vert \gamma_\bk \vert \left[\frac{1}{\lambda_{\bk,1}}
- \frac{1}{\lambda_{\bk,2}}\right] \\
&-& J_\perp S(S+1).
\eea
Using the values of $\sbar$ and $\mu$ thus obtained, we calculate the gap to triplet excitations.
The dimer-N\'eel transition occurs when the triplon gap vanishes at $J_\perp=J_{\perp c}$. Using 
the above equations, we arrive at the following two results at the critical point.
(i) The value $\sbar$ at the dimer-N\'eel critical point is independent of spin and given by 
\bea
\sbar^2_c = \frac{5}{2}
+\frac{3}{2N_\perp} \sum_\bk \!{}^{'}\left[
\frac{\vert \gamma_\bk \vert -6}{\sqrt{9-3\vert \gamma_\bk \vert}}
-\frac{\vert \gamma_\bk \vert +6}{\sqrt{9+3\vert \gamma_\bk \vert}}\right]\! .
\eea
A numerical evaluation shows $\sbar_c \approx 0.872$.
(ii) We find the location of the dimer-N\'eel critical point
\bea
\frac{J_{\perp c}}{S(S+1)}\! =\! 10 \!-\! \frac{36}{N_\perp}
\!\sum_\bk {}^{'}\!\!\left[
\frac{1}{\sqrt{9\!-\!3\vert \gamma_\bk \vert}}
\!+\!\frac{1}{\sqrt{9\!+\!3\vert \gamma_\bk \vert}}\right]\!\!.
\eea
A numerical evaluation yields $J_{\perp c} \approx 1.748 S(S+1)$.
For $S=1/2$, the mean field result, $J_{\perp c}[S=1/2] \approx 1.311$, is somewhat smaller than the
QMC value. For higher spin, the mean field critical points, 
$J_{\perp c}[S=1] \approx 3.496$ and $J_{\perp c}[S=3/2] \approx 6.555$, are significantly smaller than the
corresponding QMC results. This is summarized in Table \ref{Table:hc}.
Remarkably, as shown in Fig.\ref{fig:qcp},
the scaling result $J_{\perp c} \sim S(S+1)$ from mean field theory appears to be valid even for the exact QMC 
results on the honeycomb lattice.
We have also explicitly checked that triplon
condensation of the mode with energy $\lambda_{\bk,1}$ at momentum
$\bk=(0,0)$  yields N\'eel order on the honeycomb bilayer.

\section{Beyond mean field theory: Variational analysis}

Corrections to the mean field Hamiltonian arise from triplet-triplet interactions, and coupling to higher spin objects such as quintets and heptets.
As a function of S, we find two regimes where two different correction terms dominate. For $S = 1/2$, the only correction stems from triplet-triplet interactions since higher spin states are absent.
For $S > 1/2$, the dominant correction arises from coupling to higher spin (quintet) states.
 Ordinarily, such quintet terms can be ignored as the energy cost of exciting quintets is large; however, these terms scale as $S^2$ as opposed to the $S^0$ scaling of the
triplet-triplet interactions and they play an increasingly important role for larger $S$.
These two correction terms are separately discussed in the following two sections. 

Specifically, for the two regimes $S=1/2$ and $S > 1/2$, we identify the leading correction term and take it into account using a variational 
approach. With the leading correction, the Hamiltonian takes the form
$H^{(0)}_{_\Box} \to H^{(0)}_{_\Box} + \Delta H_{_\Box}$ and
$H^{(0)}_{\hexagon} \to H^{(0)}_{\hexagon} + \Delta H_{\hexagon}$.
We treat $\Delta H$ as a perturbation acting upon the states of $H^{(0)}$. As a variational ansatz, we
assume that the effect of the correction terms is entirely accounted for by
a renormalization of the parameters $\sbar$ and $\mu$ which enter the mean field
Hamiltonian, $H^{(0)}_{_\Box}$ or $H^{(0)}_{\hexagon}$.
We choose $\mu$ to enforce single boson occupancy per site on average. The perturbations $\Delta H$, for both regimes, preserve 
total boson number. Thus, it suffices to evaluate total boson number using $H^{(0)}$. This gives us the constraint
\bea
\sbar^2 + \sum_{i,m} \la t_{i,m}^\dg t_{i,m} \ra = N_\perp,
\eea
where the expectation value is evaluated with respect to $H^{(0)}$. (For the honeycomb lattice case, there is an additional sum over the sublattice degree of freedom in the above equation). This leads precisely to the mean field number constraint in Eq.~\ref{mftconstraint} or Eq.~\ref{Eq.hcMFTmu}, which can now be used to determine
$\mu$.
The parameter $\sbar$ is chosen to minimize the ground state energy, evaluated to leading order in perturbation 
theory. For $S=1/2$, we find that the leading correction is obtained within first order perturbation theory in
$\Delta H$. For $S > 1/2$, the dominant perturbing terms require us to go to second order in perturbation theory.
In the next two sections, we discuss these correction terms in detail. 

\section{Triplet interaction corrections}

\subsection{Triplet Interactions on square lattice}

Staying within the singlet-triplet sector, the term we have ignored in the mean field treatment is the
triplet-triplet interaction term. For $S=1/2$, there are no higher spin bosons beyond the singlet-triplet sector,
so this is the only correction. For $S > 1/2$, this constitutes one term in a slew of correction terms.
For any spin $S$, the triplet interaction terms are given by 
\bea
\!\Delta H_{\Box}^{\rm (t)} \! =\! -\frac{1}{2}\!\sum_{\la ij \ra}\!\!\!
\sum_{\begin{array}{c}
u,v,w,v',w' \\
\in\{x,y,z\}\end{array}
}\!\!\!\!\!\!\!\!\!\!\!\!
\epsilon_{uvw} t_{i,v}^\dg t_{i,w}\epsilon_{uv'w'}t_{j,v'}^\dg t_{j,w'}\!.
\label{eq.tripint}
\eea
We note that there are no cubic terms in triplet operators. As described in Ref.~\onlinecite{vojta2011}, this makes our bilayer problem qualitatively different from other dimerized states such as the spin-1/2 staggered dimer on the square lattice. 
Typically, triplet-triplet interactions such as those of Eq. \ref{eq.tripint} are taken into account within a self-consistent Hartree-Fock approximation.\cite{sachdevbhatt1990,gopalan1994}
Here, we take the interactions in Eq. \ref{eq.tripint} to be a perturbation acting on $H_\Box^{(0)}$ and 
evaluate the first order correction to ground state energy. To this end, we decouple $\Delta H_\Box^{\rm (t)}$ using bilinears that possess finite expectation values at the level of mean field theory:
\bea
\label{rhodeltasq}
\nn \la t_{i,v}^\dg t_{i+\bdelta,w} \ra \equiv \delta_{v,w} \rho, \\
\la t_{i,v}^\dg t_{i+\bdelta,w}^\dg \ra \equiv \delta_{v,w} \Delta.
\eea
Here, $i$ and $i+\bdelta$ are nearest neighbours on the square lattice. Explicit expressions for $\rho$ and $\Delta$ are given in Appendix \ref{app.sqBog}. We note that $\rho$ and $\Delta$ are functions of the variational parameters $\sbar$ and $\mu$.
The first order energy correction due to triplet interactions is given by 
\bea
\Delta E_\Box^{\rm (t)} =\la \Delta H_\Box^{\rm (t)} \ra =  6 N_\perp \left[ \rho^2-\Delta^2 \right].
\eea
Thus, the variational energy of the ground state upon including the triplet interaction
term is given by
\bea
E_{\Box,\rm var}^{{\rm (t)}} (\sbar,\mu)= E_{\Box}^{(0)}+\Delta E_\Box^{{\rm (t)}},
\eea
where $E_\Box^{(0)}$ is as defined in Eq.~\ref{E2GS}.
The parameter $\sbar$ is chosen to minimize this energy. We find that the triplet interactions reduce
the stability of the dimer phase and shift $J_{\perp c}$ to larger values.
For $S=1/2$, this leads to a renormalized transition point $J_{\perp c} \approx 2.58$, very close to
the QMC result. For $S > 1/2$, the renormalization is too weak to account for the discrepancy between
the earlier mean field result and the QMC data. These triplet corrected results for the square lattice are
summarized in Table \ref{Table:square}.

\subsection{Honeycomb}
The interaction between triplets on the honeycomb lattice is given by
\bea
\nn \Delta H_{\hexagon}^{\rm (t)} = -\frac{1}{2}\!\sum_{i,\bdelta}\!\!\!\sum_{
\begin{array}{c}
u,v,w,v',w'\\
\in\{x,y,z\}\end{array}
}\!\!\!\!\!\!\!
\epsilon_{uvw} t_{i,A,v}^\dg t_{i,A,w} \times \\
\epsilon_{uv'w'}t_{i+\bdelta,B,v'}^\dg t_{i+\bdelta,B,w'}.
\eea
The operators $\bdelta$ are such that the sites $(i,A)$ and $(i+\bdelta,B)$ are nearest neighbours.
This interaction term contributes to the ground state energy at first order in perturbation theory. To evaluate this correction, we quadratically decompose the interaction using the following two bilinears:
\bea
\nn \la t_{i,A,v}^\dg t_{i+\bdelta,B,w} \ra \equiv \delta_{v,w} \rho, \\
\label{rhodeltahc}
\la t_{i,A,v}^\dg t_{i+\bdelta,B,w}^\dg \ra \equiv \delta_{v,w} \Delta,
\eea
with the expectation values to be evaluated using the unperturbed Hamiltonian $H_{\hexagon}^{(0)}$. Explicit expressions for $\rho$ and $\Delta$ are given in Appendix \ref{app.hcBog}.
The first order correction to ground state energy is given by 
\bea
\Delta E^{\rm (t)}_{\hexagon}=\frac{9}{2} N_\perp [\rho^2 - \Delta^2].
\eea
The parameter $\sbar$ is chosen to minimize the energy 
\bea
E_{\hexagon,\rm var}^{\rm (t)}(\sbar,\mu)= E_{\hexagon}^{(0)}+\Delta E_{\hexagon}^{\rm (t)}.
\eea 
As on the square lattice, we find that the triplet interactions reduce
the stability of the dimer phase and shift $J_{\perp c}$ to larger values.
For $S=1/2$, this leads to a renormalized transition point $J_{\perp c} \approx 1.59$, which is in reasonable
agreement with the QMC result $J_{\perp c} = 1.645(1)$. For $S > 1/2$, however, the renormalization is 
again too weak to account for the QMC data. These triplet corrected results for the critical point on the
honeycomb lattice are
summarized in Table \ref{Table:hc}.

\section{Quintet corrections}

In the previous section, we have seen that triplet correction terms lead to a reasonably good agreement
with QMC results for the dimer-N\'eel quantum critical point for $S=1/2$. However, they fail to account for the significant discrepancy
between QMC and bond operator mean field theory for $S > 1/2$. This leads to us to suspect that higher order 
spin excitations on the dimer bonds must be responsible for this difference.
Upon including quintet terms, the spin operators at a site $i$ contain a large number of terms as given in
Eq.~(20) and Eq.~(21) of Ref.~\onlinecite{brijesh2010} and reproduced in Appendix \ref{app.spinquintet} for convenience.

Using these spin expressions to rewrite the Hamiltonian in Eq.~\ref{genHamiltonian},
we find that correction terms beyond mean field theory, including those involving quintet states, may be grouped as
\bea
\Delta H = \! \hat{D}_{tttt}\!+\! \sbar \hat{R}_{ttq}(S^2) \!+\! \hat{F}_{ttqq}(S^2) \!+\!
\hat{G}_{qqqq}(\! S^0\!).
\label{inplaneexchange}
\eea
The subscripts indicate the composition of the terms in terms of bond operators. The scaling of each term with $S$ 
is indicated in parentheses. For example, $\hat{R}_{ttq}(S^2)$ is composed of terms which involve two triplet 
operators and one quintet operator, and the coefficients of these terms scale as $S^2$. The term which we have
accounted for in the previous section is $\hat{D}_{tttt}$, which scales as $S^0$ and contains four triplet operators.
Na\"ively, terms involving quintets should be less important due to the energy cost of exciting quintets. However, we see 
from the above classification of terms that the coefficients of $\hat{R}_{ttq}(S^2)$ and $\hat{F}_{ttqq}(S^2)$
increase rapidly with increasing spin. We find 
that $\hat{R}_{ttq}(S^2)$ is, in fact, the dominant contribution for all $S > 1/2$.
(For the case of $S=1$, we have explicitly checked that this term dominates over triplet-triplet interactions 
encoded by $\hat{D}_{tttt}$ - see Table \ref{Table:square}). The term $\hat{F}_{ttqq}(S^2)$ is suppressed because it involves two quintet operators which act on different sites. In our variational scheme, this term will contribute to the ground state energy at second order in perturbation theory. However, as the quintets are taken to be localized excitations, this term will involve intermediate states
with two quintet excitations. Therefore, it will contribute much less than $\hat{R}_{ttq}(S^2)$.

In the vicinity of the dimer-N\'eel transition, we assert that $\hat{R}_{ttq}(S^2)$ will remain the dominant correction term for any $S>1/2$ even if higher spin states such as heptets, nonets, etc., are included. As the dimer-N\'eel transition occurs via condensation of triplet excitations, it is reasonable that the dominant corrections come from quintets which are immediately higher in energy than triplets. Heptets, nonets, etc. occur at much higher energies and are unlikely to affect the triplet condensation point. To argue that this is indeed the case, we first note that the Hamiltonian of Eq.\ref{genHamiltonian} can change the spin of a bond by $\pm 1$ at most (this can be seen from the rotation properties of a single spin operator acting on a bond eigenstate).
For example, if we restrict our attention to one particular bond, the Hamiltonian connects a triplet state to singlet, triplet and quintet states. The matrix element connecting the triplet to a nonet state (or a state of even higher spin) is zero. Similarly, on a given bond, the heptet state has non-zero matrix elements only with quintet, heptet and nonet states. 
The resulting terms in the Hamiltonian involving heptets, nonets, etc. will \textit{not} contribute at second order in perturbation theory, but will only appear at higher order. As an illustration, upon including heptets, the Hamiltonian can have a term of the form $h_{i,m}^\dg q_{i,n} t_{j,m'}^\dg t_{j,n'}$. Clearly, this term does not contribute to ground state energy at first or second order. In addition, at whichever order it contributes, the energy denominators will involve large heptet excitation energies which will further suppress the energy contribution. 

In summary, in the vicinity of the dimer-N\'eel transition for any value of $S>1/2$, the leading correction to bond operator mean field theory comes from $\sbar\hat{R}_{ttq}(S^2)$. We write
\bea
\Delta H^{(S > 1/2)}  \approx  \Delta H^{\rm (q)} \equiv \sbar\hat{R}_{ttq}(S^2).
\eea
Note that if we were to use a path integral approach to integrate out the quintet excitations at this stage, we 
would be led to an {\it effective} triplet-triplet interaction which is enhanced by a factor of $S^4$ compared to the
bare triplet-triplet term discussed in the previous section (although it would be divided by the quintet
gap which scales as $S^2$ near the N\'eel to dimer transition). Here we follow a different route, similar in spirit, and
treat this term perturbatively assuming the quintet states to be local excitations. The energy cost of 
creating a quintet is given by 
Eq.~\ref{ILexch}. We measure this energy cost from the Lagrange multiplier $\mu$, to get 
\bea
\varepsilon_{q}-\mu = J_\perp\{3-S(S+1)\}-\mu
\eea
as the energy cost of a quintet excitation.

\subsection{Quintet corrections on the square lattice}
\label{sec.sqqnt}

The terms in $\sbar\hat{R}_{ttq}(S^2)$ may be organized as 
\bea
\sbar \hat{R}_{ttq}(S^2) = 
\sbar\sum_{i}\sum_{n=-2,\cdots,2}\left[ q_{i,n}^\dg \sum_{\bdelta}\hat{T}_{i,i+\bdelta}^{[n]} + h.c.\right].
\eea
The operator $\hat{T}_{i,i+\bdelta}^{[n]}$ is composed of triplet bilinears. The index $\bdelta$ sums over the four nearest neighbour vectors on the square lattice. The explicit form of these operators is given in Appendix \ref{app.sqquintet}.
The operator $\hat{R}_{ttq}(S^2)$ does not contribute to ground state energy at first order, as it is linear in quintet operators. The energy correction, at second order, is given by
\bea
\Delta E_\Box^{\rm (q)} \!\!=\sbar^2 \!\!\sum_{\sigma\neq 0}\!\!\!\!\! \sum_{
\begin{array}{c}
i,n,\bdelta \\
i',n',\bdelta'
\end{array}}\!\!\!\!\!\!\!\!
%i,i',n,n',\bdelta,\bdelta'}\!\!\!\!
\frac{\la 0 \vert
q_{i',n'} (\hat{T}^{[n]}_{i',i'+\bdelta'})^\dg
\vert \sigma \ra \la \sigma\vert
 q_{i,n}^\dg \hat{T}_{i,i+\bdelta}^{[n]}
\vert 0 \ra }{E_0-E_\sigma}.
\eea
The index $\sigma$ sums over all excited states of $H_{\Box}^{(0)}$, the variational Hamiltonian. The only intermediate states that contribute are those with a single quintet.
In our variational formalism, we take the quintets to be local excitations. This constrains us to $(i=i')$, $(n=n')$. This leaves us with
\bea
\!\Delta E_\Box^{\rm (q)} \!=\!\! \sbar^2\!\!\sum_{\nu\neq 0}\! \sum_{i,n}\!\!
\frac{\la 0 \vert
\!\!\sum_{\bdelta'}(\hat{T}^{[n]}_{i,i+\bdelta'})^\dg
\vert \nu \ra \la \nu\vert
\!\!\sum_{\bdelta} \hat{T}_{i,i+\bdelta}^{[n]}
\vert 0 \ra }{E_0-E_\nu}.
\label{Equintsq}
\eea
The intermediate states $\vert \nu\ra$ which contribute involve a single quintet excitation. Within the triplet sector, at zero temperature, the intermediate states can have either (a) no triplon quasiparticles, or (b) two triplon quasiparticles. The contribution from states with no triplon quasiparticles vanishes due to global spin-rotational symmetry of the Hamiltonian. 
The energy correction from two triplon intermediate states is evaluated 
to obtain the energy correction, $\Delta E_\Box^{\rm (q)}$. The complete expression is given in Appendix 
\ref{app.sqquintet}.

Being second order in $\hat{R}_{ttq}(S^2)$, the energy correction from quintet coupling na\"ively scales as $S^4$ for
large $S$. However, the energy denominator involves the energy of quintet states which is proportional to $J_\perp$. Close to the dimer-N\'eel transition, at mean field level, $J_\perp$ approximately scales as $S^2$ for large $S$
(see Eq.~\ref{jccrit}). We expect perturbative corrections to preserve this scaling of $J_{\perp c}$ with $S^2$. 
Thus, near the dimer-N\'eel transition, $\Delta E_\Box^{\rm (q)}$ scales as $S^4/S^2\sim S^2$.
The ground state energy to leading order in perturbation theory is thus given by 
\bea
E_{\Box,\rm var}^{(S > 1/2)} (\sbar,\mu)= E_{\Box}^{(0)} + \Delta E_\Box^{\rm (q)}.
\eea
This energy is a function of $\sbar$ and $\mu$. As discussed earlier, $\mu$ is tuned to enforce single boson occupancy per site, while $\sbar$ is chosen to minimize $E_{\Box,\rm var}^{(S > 1/2)} (\sbar,\mu)$.

Having determined $\sbar$ and $\mu$ variationally, we can find the gap to triplon excitations as a function of $J_\perp$. The Dimer-N\'eel transition is indicated by the vanishing of the triplon gap in the variationally obtained state. As summarized
in Table \ref{Table:square}, the renormalized critical points obtained in this manner are within $2.5\%$ of the QMC results.
While the precise quantitative agreement is perhaps fortuitous, and will certainly change depending on the nature of
the approximations made, the important problem we have resolved is to show that the large discrepancy between QMC
and simple bond operator mean field theory for $S > 1/2$ can be accounted for by virtual quintet excitations.

\begin{table}
%\caption{Value of $J_{\perp c}$ on the square lattice.}
\begin{tabular}{|c||c||c|c|c|}
\hline
S & QMC & MFT & MFT + & MFT +\\
 & & & triplet interactions  & quintet coupling \\
\hline
\hline
0.5 & 2.5220(1) \cite{sandvik1994} & 2.287 & 2.568 & - \\
1 & 7.150(2) & 6.098 & 6.387 & 7.32(14)  \\
1.5 & 13.634(3) &  11.434 & 11.714 & 13.32(1) \\%quint data from quintS1p5L90b.dat
\hline
\end{tabular}
\caption{Value of $J_{\perp c}$ on the square lattice from different methods for different values of S. MFT stands for mean field Theory.
The QMC data for S=1/2 is from Ref.~\onlinecite{sandvik1994}. The column `MFT+Triplet interactions' gives variational results appropriate for S=1/2. The column `MFT+quintet coupling' gives variational results appropriate for $\mathrm{S > 1/2}$.}
\label{Table:square}
\end{table}

\subsection{Quintet corrections on the honeycomb lattice}

On the honeycomb lattice, the terms in $\sbar\hat{R}_{ttq}(S^2)$  may be written as
\bea
\nn \sbar\hat{R}_{ttq} = \sbar\sum_i \sum_{n=-2,\cdots,2}\left[
q_{i,A,n}^\dg \sum_\bdelta \hat{A}_{i,i+\bdelta}^{[n]}+h.c \right.\\
\left. +q_{i,B,n}^\dg \sum_\bdelta \hat{B}_{i,i-\bdelta}^{[n]} + h.c.\right].
\eea
The operators $\hat{A}_{i,i+\bdelta}$ and $\hat{B}_{i,i-\bdelta}$ are triplet bilinears centred on nearest neighbour bonds. We give their explicit forms in momentum space in Appendix \ref{app.hcquintet}.
The terms in $\hat{R}_{ttq}(S^2)$ contribute to ground state energy only at second order in perturbation theory. The energy correction may be written as
\bea
\nn \Delta E_{\hexagon}^{\rm (q)} =\sbar^2 \sum_{\sigma\neq 0} 
\sum_{i,n}
\la 0 \vert \left[
q_{i,A,n} \sum_{\bdelta'} \hat{A}_{i,i+\bdelta'}^{[n]}\right]
\vert \sigma \ra \times \\
\la \sigma \vert \left[
q_{i,A,n}^\dg \sum_\bdelta \hat{A}_{i,i+\bdelta}^{[n]}
\right]\vert 0 \ra
/\{E_{0}-E_{\sigma}\} + (A\rightarrow B),
\eea
where the index $\sigma$ sums over all excited states of $H_{\hexagon}^{(0)}$. As the terms in $\hat{R}_{ttq}(S^2)$ involve one quintet operator, only intermediate states with a single occupied quintet state will contribute. 
\bea
\nn \Delta  E_{\hexagon}^{\rm (q)} = \sbar^2\sum_{\nu\neq 0} 
\sum_{i,n}
\frac{
\la 0 \vert \sum_{\bdelta'} (\hat{A}_{i,i+\bdelta'}^{[n]})^\dg
\vert \nu \ra 
\la \nu \vert 
\sum_\bdelta \hat{A}_{i,i+\bdelta}^{[n]}
\vert 0 \ra
}{\{E_{0}-E_{\nu}\}}\\
 + \sbar^2\sum_{\nu\neq 0} 
\sum_{i,n}\!\!
\frac{
\la 0 \vert \sum_{\bdelta'} (\hat{B}_{i,i-\bdelta'}^{[n]})^\dg
\vert \nu \ra 
\la \nu \vert 
\sum_\bdelta \hat{B}_{i,i-\bdelta}^{[n]}
\vert 0 \ra
}{\{E_{0}-E_{\nu}\}}.\phantom{ab}
\label{hc
Equint}
\eea

We evaluate these overlaps in momentum space, as described in Appendix 
\ref{app.hcquintet}. The intermediate state $\vert\nu\ra$ could have either (i) no triplon quasiparticles, or (ii) two triplon quasiparticles.
However, the contribution from states with no triplons vanishes due to global spin rotational symmetry.
The explicit expression for $\Delta E_{\hexagon}^{\rm (q)}$ is given in Appendix \ref{app.hcquintet}.

Thus, the energy of the ground state to leading order in quintet coupling, is given by
\bea
E_{\hexagon,\rm var}^{(S>1/2)} = E_{\hexagon}^{(0)} + \Delta E_{\hexagon}^{\rm (q)}.
\eea
We choose $\sbar$ to minimize this energy. The vanishing of the triplet gap in the variationally determined state signals the dimer-N\'eel transition. Our results for $J_{\perp c}$ on the honeycomb lattice are shown in Table \ref{Table:hc}. The
renormalized critical points for $S=1, 3/2$ are within $5\%$ of the QMC value. 

\begin{table}
%\caption{Value of $J_{\perp c}$ on the honeycomb lattice.}
\begin{tabular}{|c||c||c|c|c|}
\hline
S & QMC & MFT & MFT + & MFT +\\
 & & & triplet interactions & quintet coupling \\
\hline
\hline
0.5 & 1.645(1) & 1.312 & 1.588 & - \\
1 & 4.785(1) & 3.498 & 3.774 & 4.80(9) \\
1.5 & 9.194(3) & 6.559 & 6.837 & 9.58(18) \\
\hline
\end{tabular}
\caption{Value of $J_{\perp c}$ on the honeycomb lattice from different methods for different values of S. MFT stands for mean field Theory.}
\label{Table:hc}
\end{table}

\section{Discussion}
In this paper, we have studied the N\'eel to dimer transition in Heisenberg antiferromagnets on bilayer square and honeycomb
lattices for different spin values using QMC and bond operator approaches. 
The critical bilayer exchange $J_{\perp c}$ scales as $S(S+1)$ within, both, bond operator mean field theory and QMC simulations. However, there is a systematic deviation between bond operator mean field theory and QMC, with the deviation itself scaling as $\sim S^2$. Our variational extension of bond operator theory to include the dominant
triplet and quintet excitations successfully captures this systematic deviation and gives a more precise estimate of 
$J_{\perp c}$.

Bi$_3$Mn$_4$O$_{12}$(NO$_{3}$) provides an example of a bilayer honeycomb antiferromagnet\cite{BMNO} with
$S=3/2$, where strong interlayer exchange couplings $\sim 2 J_1$ have been inferred from electronic structure
calculations\cite{vandenBrink}. Despite this strong bilayer coupling, our study indicates that this material would be deep in the N\'eel ordered phase if there are no other frustrating interactions.
We are thus forced to attribute the observed lack of magnetic order in Bi$_3$Mn$_4$O$_{12}$(NO$_{3}$) to
frustration effects arising from further neighbor couplings; such further neighbor interactions have been shown to drive
a variety of new phases on the honeycomb lattice.\cite{Jafari2011, albuquerque2011, FaWang2010, YingRan2011, Clark2010, Kawamura, Mulder2010, Ganesh2011}
One recent example of a dimer system with $S=1$ is the triangular dimer material\cite{Stone2008,Samulon2009} Ba$_3$Mn$_2$O$_8$.
Our approach could be applied to understand the triplon spectrum and the effect of quintet corrections in this
material.
In particular, our work shows that extracting exchange couplings from fitting experimental data to bond operator mean field theory may not yield precise estimates.
In summary, our work provides a starting point to think about the
physics of high spin Heisenberg antiferromagnets in a variety of model systems and materials.

\acknowledgments

This research was
supported by the Canadian NSERC (RG, AP), an Ontario Early Researcher Award (RG, AP),
and the Swiss HP$^2$C initiative (SVI).
The quantum Monte Carlo simulations were performed on the Brutus cluster at ETH Zurich.
RG thanks J. Rau, V. Vijay Shankar and C. M. Puetter for discussions. 

\appendix
\section{Square bilayer: bosonic Bogoliubov transformation}
\label{app.sqBog}
The MFT Hamiltonian of Eq.~\ref{MFTHamiltonian} is diagonalized by a pseudounitary matrix,
\bea
U_\bk=\left(\begin{array}{cc}
\cosh\theta_\bk & \sinh\theta_\bk\\
\sinh\theta_\bk & \cosh\theta_\bk
\end{array}\right).
\eea
Imposing $\tanh 2 \theta_\bk = -2\epsilon_\bk/(A+2 \epsilon_\bk)$, we get
\bea
\psi_{\bk,u}^\dg \!\! \left(\!\!\begin{array}{cc}
A+2\epsilon_\bk & 2\epsilon_\bk \\
2\epsilon_\bk & A+2\epsilon_\bk \end{array}\!\!\right)
\!\!\psi_{\bk,u} = \phi_{\bk,u}^\dg\!\!\left(\!\!\begin{array}{cc}
\lambda_\bk & 0 \\
0 & \lambda_\bk \end{array}\!\!\right)\!\!\phi_{\bk,u}.
\eea
We have defined new quasiparticle operators given by $\psi_{\bk,u}=U_\bk \phi_{\bk,u}$ so that
\bea
\left(\begin{array}{c}
t_{\bk,u} \\ t_{-\bk,u}^\dg \end{array}\right)\!\!
=\!\left( \begin{array}{cc}
\cosh \theta_\bk & \sinh \theta_\bk\\
\sinh\theta_\bk & \cosh \theta_\bk
\end{array}\right)
\left(\begin{array}{c}
\tau_{\bk,u} \\ \tau_{-\bk,u}^\dg \end{array}\!\!\right).
\label{sqBogoliubov}
\eea
The $\tau$ operators are the triplon quasiparticles. The bilinears defined in Eq.~\ref{rhodeltasq}, may be evaluated using the elements of U as follows:
\bea
\nn \rho &=& \frac{1}{4N_\perp}\sum_{\bk, \bdelta}\left[\la t_{\bk,v}^\dg t_{\bk,v}\ra e^{i\bk.\bdelta}\right]\\
&=&\frac{1}{4N_\perp}\sum_{\bk}{}^{'}(2\cos k_x + 2\cos k_y)\frac{A+2\epsilon_\bk}{\lambda_\bk}, \\
\nn \Delta &=& \frac{1}{4N_\perp}\sum_{\bk,\bdelta}\left[\la t_{\bk,v}^\dg t_{-\bk,v}^\dg\ra e^{i\bk.\bdelta}\right] \\
&=&\frac{1}{4N_\perp}\sum_{\bk}{}^{'}(2\cos k_x + 2\cos k_y)\frac{(-2\epsilon_\bk)}{\lambda_\bk}. 
\eea

\section{Honeycomb bilayer: bosonic Bogoliubov transformation}
\label{app.hcBog}
The mean field Hamiltonian of Eq.~\ref{hcmfthamiltonian} may be diagonalized by the matrix,
\bea
\nn
P_{\bk}= \frac{1}{\sqrt{2}}\!\!\left(\!\!\!\begin{array}{cccc}
1 & 1 & 0  & 0 \\
-b_\bk & b_\bk & 0 & 0 \\
0 & 0 & 1  &  1 \\
0 & 0 & -b_\bk & b_\bk
\end{array}\!\!\right)\!\!\!
\left(\!\!\begin{array}{cccc}
C_{\bk,1}\!\!\! & 0\!\!\! & S_{\bk,1}\!\!\! & 0\! \\
0\!\!\! & C_{\bk,2}\!\!\! & 0\!\!\! & S_{\bk,2}\! \\
S_{\bk,1}\!\!\! & 0\!\!\! & C_{\bk,1}\!\!\! & 0\! \\
0\!\!\! & S_{\bk,2}\!\!\! & 0\!\!\! & C_{\bk,2}\!
\end{array}\!\right). 
\eea
Here, we have defined $b_\bk \equiv \beta_\bk^*/\vert \beta_\bk \vert$. We take the other entries to be hyperbolic functions given by $C_{\bk,n}=\cosh\kappa_{\bk,n}$ and $S_{\bk,n}=\sinh\kappa_{\bk,n}$, with $n=1,2$.
With this definition,
this matrix $P_\bk$ satisfies the pseudo-unitarity condition $P_\bk\sigma P_\bk^\dg=\sigma$, where $\sigma=\mathrm{Diag}\{1,1,-1,-1\}$. 
To diagonalize the Hamiltonian matrix $M_\bk$, we set 
\bea
\nn \tanh 2\kappa_{\bk,1} &=& \beta_\bk/(C-\beta_\bk);\\
\tanh 2\kappa_{\bk,2} &=& -\beta_\bk/(C+\beta_\bk).
\eea
With this choice, the matrix $P_\bk$ diagonalizes the Hamiltonian,  
\bea
P_\bk^\dg M_\bk P_\bk = \mathrm{Diag}\{\lambda_{\bk,1},
\lambda_{\bk,2},\lambda_{\bk,1},\lambda_{\bk,2}  \}.
\eea
where $\lambda_{\bk,1/2}$ are as defined in the main body.
We transform the triplet operators defined in 
Eq.~\ref{Eq.hcmftpsiM} into new quasiparticle operators using
\bea
\left(
\begin{array}{c}
t_{\bk,A,u} \\
t_{\bk,B,u} \\
t_{-\bk,A,u}^\dg \\
t_{-\bk,B,u}^\dg\end{array}\right)
= P_{\bk}\left(
\begin{array}{c}
\vartheta_{\bk,1,u} \\
\vartheta_{\bk,2,u} \\
\vartheta_{-\bk,1,u}^\dg \\
\vartheta_{-\bk,2,u}^\dg\end{array}\right).
\eea
The $\vartheta$ operators are the triplon quasiparticles. Compared to the square lattice case, the quasiparticle operators have an additional index on account of the sublattice degree of freedom. 
We can express our original triplet operators as follows:
\bea
\nn t_{\bk,A,u}&=&\sum_{f=1,2}\left( C_{\bk,f}
\vartheta_{\bk,f,u} + S_{\bk,f}\vartheta_{-\bk,f,u}^\dg  \right), \\
t_{-\bk,B,u}\!\!\!&=&\!\!\!\sum_{f=1,2}\!\!(-1)^{f} b_{\bk}^* \left(\! 
C_{\bk,f}\vartheta_{-\bk,f,u} \!+\! S_{\bk,f}\vartheta_{\bk,f,u}^\dg \!\right)\!\!.
\label{Eq.tripquasiparticles}
\eea
The bilinears defined in Eq.~\ref{rhodeltahc} can be evaluated as 
\bea
\nn \rho&=&\frac{2}{3N_\perp}\sum_{\bk}{}
\la t_{\bk,A,v}^\dg t_{\bk,B,v}\ra \gamma_\bk \\
&=&\frac{1}{6N_\perp}\sum_{\bk}
\vert \gamma_\bk\vert\left[
-\frac{C-\vert \beta_\bk \vert}{\lambda_{\bk,1}} + \frac{C+\vert \beta_\bk \vert}{\lambda_{\bk,2}}
\right],\\
\nn \Delta&=&\frac{2}{3N_\perp}\sum_{\bk}
\la t_{\bk,A,v}^\dg t_{-\bk,B,v}^\dg \ra \gamma_\bk \\
&=&\frac{-1}{6N_\perp}
\sum_{\bk} \vert \gamma_\bk\vert
\left[\frac{\vert \beta_\bk \vert}{\lambda_{\bk,1}} + \frac{\vert \beta_\bk \vert}{\lambda_{\bk,2}}\right].
\eea

\section{Spin operator expressions including quintet terms}
\label{app.spinquintet}
Including triplet and quintet operators, the complete expression for the spin operators on the two layers
of the bilayer are \cite{brijesh2010}
%\begin{widetext}
\bea
S^{+}_{i,\ell=1,2} &=& (-1)^\ell \sqrt{\frac{2S(S+1)}{3}} \sbar ( t_{i,-1} -t^\dg_{i,1}  ) \nn  \\
&+& (-1)^\ell \sqrt{\frac{(2S-1)(2S+3)}{5}} \big[ ( t^\dg_{i,-1} q_{i,-2}-q^\dg_{i,2} t_{i,1}  ) \nn \\
&+& \sqrt{\frac{1}{2}}
( t^\dg_{i,0} q_{i,-1}-q^\dg_{i,1} t_{i,0} )
 +\sqrt{\frac{1}{6}} ( t_{i,1}^\dg q_{i,0}-q_{i,0}^\dg t_{i,-1} ) \big] \nn \\
&+& \sqrt{\frac{1}{2}} (t^\dg_{i,1} t_{i,0}+t^\dg_{i,0} t_{i,-1})
+\sqrt{\frac{3}{2}} (q^\dg_{i,1} q_{i,0} + q^\dg_{i,0} q _{i,-1}) \nn \\
&+& q^\dg_{i,2} q_{i,1} + q^\dg_{i,-1} q_{i,-2},
\eea
with $S^{-}_{i,\ell=1,2}$ being its Hermitian conjugate, while
\bea
S_{i,\ell=1,2}^z  &=&  (-1)^\ell \sqrt{\frac{S(S+1)}{3}}
\sbar ( t_{i,0}+t^\dg_{i,0} )  \nn  \\
&+& (-1)^\ell \sqrt{\frac{(2S-1)(2S+3)}{5}} \big[ \sqrt{\frac{1}{3}}
( t^\dg_{i,0} q_{i,0} \!+\! q^\dg_{i,0} t_{i,0} ) \nn \\
&+& \frac{1}{2} ( t^\dg_{i,1} q_{i,1} + q^\dg_{i,1} t_{i,1}
+t_{i,-1}^\dg q_{i,-1} + q^\dg_{i,-1} t_{i,-1} ) \big]  \nn \\
&+& \frac{1}{2}  ( t^\dg_{i,1} t_{i,1} - t^\dg_{i,-1} t_{i,-1} 
+ q^\dg_{i,1} q_{i,1} - q^\dg_{i,-1} q_{i,-1} ) \nn \\
&+& q^\dg_{i,2} q_{i,2}-q^\dg_{i,-2} q_{i,-2}.
\eea
%\end{widetext}

\section{Square Bilayer: inclusion of quintets}
\label{app.sqquintet}
The spin operators with the inclusion of quintets are given in Eq.~21 of Ref.~\onlinecite{brijesh2010}. Using this reference, we now give explicit expressions for $\hat{R}_{ttq}(S^2)$. In the main text, we defined $\hat{R}_{ttq}(S^2)$ in terms of triplet bilinears  $\hat{T}_{i,i+\delta}^{[n]}$. 
Here, we give expressions for $\hat{T}_{i,i+\delta}^{[n]}$ in momentum space. We use the Fourier transform convention 
\bea
t_{i,u\in\{x,y,z\}}=\frac{1}{\sqrt{N_\perp}}\sum_\bk
t_{\bk,u}e^{i\bk.\br_i}.
\eea
The operator $\hat{T}_{i,i+\delta}^{[n]}$ is composed of bilinears of the form $t_{i,u}(t_{i+\bdelta,v}\pm t_{i+\bdelta,v}^\dg)$. Using the Fourier transform, this generic bilinear may be written as $(1/N_\perp)\sum_{\bk,\bp}t_{-\bk+\bp,u}(t_{\bk,u}\pm t_{-\bk,u}^\dg)e^{i\bk.\bdelta}e^{i\bp.\br_i}$.

Thus, we may write
\bea
\label{Tmomentumspace}
\sum_{\bdelta}\hat{T}_{i,i+\bdelta}^{[n]} = \frac{M}{N_\perp}
\sum_{\bk,\bp} \hat{T}_{-\bk+\bp,\bk}^{[n]}e^{i\bp.\br_i} \eta_\bk,
\eea
where $\eta_\bk = \sum_\bdelta e^{i\bk\cdot\bdelta} = 2(\cos k_x + \cos k_y)$ and the coefficient $M=\sqrt{\frac{S(S+1)(2S-1)(2S+3)}{30}}$. The explicit forms of $\hat{T}_{-\bk+\bp,\bk}^{[n]}$ are:
\bea
\nn \hat{T}_{-\bk+\bp,\bk}^{[-2]}\!\!&=&\!\! \tilde{t}_{-\bk+\bp,x}(t_{\bk,x}+t_{-\bk,x}^\dg)
-\tilde{t}_{-\bk+\bp,y}(t_{\bk,y}\!+\! t_{-\bk,y}^\dg)\\
\nn &+&\!\! i\tilde{t}_{-\bk+\bp,x}(t_{\bk,y}\!+\! t_{-\bk,y}^\dg)
+i\tilde{t}_{-\bk+\bp,y}(t_{\bk,x}\!+\! t_{-\bk,x}^\dg)\\
\nn \hat{T}_{-\bk+\bp,\bk}^{[-1]}\!\! &=& \!\!
\tilde{t}_{-\bk+\bp,z}(t_{\bk,x}\!+\! t_{-\bk,x}^\dg)
+\tilde{t}_{-\bk+\bp,x}(t_{\bk,z}\!+\! t_{-\bk,z}^\dg)\\
\nn &+&i\tilde{t}_{-\bk+\bp,z}(t_{\bk,y}\!+\! t_{-\bk,y}^\dg)
+i\tilde{t}_{-\bk+\bp,y}(t_{\bk,z}\!+\! t_{-\bk,z}^\dg)\\
\nn \hat{T}_{-\bk+\bp,\bk}^{[0]}\!\! &=&\!\!
\sqrt{\frac{2}{3}}\left[ -\tilde{t}_{-\bk+\bp,x}(t_{\bk,x}\!+\!t_{-\bk,x}^\dg) \right. \\
\nn &-&\!\! \left. \tilde{t}_{-\bk+\bp,y}(t_{\bk,y}\!+\! t_{-\bk,y}^\dg)
\!+\! 2 \tilde{t}_{-\bk+\bp,z}(t_{\bk,z} \!+\! t_{-\bk,z}^\dg) \right] \\
\nn \hat{T}_{-\bk+\bp,\bk}^{[-1]}\!\! &=& \!\!
-\tilde{t}_{-\bk+\bp,z}(t_{\bk,x}\!+\! t_{-\bk,x}^\dg)
-\tilde{t}_{-\bk+\bp,x}(t_{\bk,z}\!+\! t_{-\bk,z}^\dg)\\
\nn &+&i\tilde{t}_{-\bk+\bp,z}(t_{\bk,y}\!+\! t_{-\bk,y}^\dg)
+i\tilde{t}_{-\bk+\bp,y}(t_{\bk,z}\!+\! t_{-\bk,z}^\dg)\\
\nn \hat{T}_{-\bk+\bp,\bk}^{[2]}\!\! &=& \!\! \tilde{t}_{-\bk+\bp,x}(t_{\bk,x}+t_{-\bk,x}^\dg)
-\tilde{t}_{-\bk+\bp,y}(t_{\bk,y}\!+\! t_{-\bk,y}^\dg)\\
\label{Texpns} -\! i\!\!\!\!\! &\phantom{a}& \!\!\!\!\!\! 
\tilde{t}_{-\bk +\bp,x}(t_{\bk,y}\!+\! t_{-\bk,y}^\dg\!)
\!\!-i \tilde{t}_{-\bk+\bp,y}(t_{\bk,x}\!+\! t_{-\bk,x}^\dg\!)
\eea

We have denoted some triplet operators as $t$ and some as $\tilde{t}$. For the purposes of the square lattice, this distinction can be ignored. 
We will use these same expressions in the context of the honeycomb lattice also. For the honeycomb case, $t$ and $\tilde{t}$ operators will act on different sublattices.

The energy correction due to coupling to quintets is given in 
Eq.~\ref{Equintsq}. Using the Fourier transformed expression in Eq.~\ref{Tmomentumspace}, we rewrite the energy as
\bea
\Delta E_\Box^{S>1/2}\!\! =\!\! \frac{M^2 \sbar^2}{N_\perp} \!\!
\sum_{m=-2,\cdots,2} \sum_{\bp} E_{\bp}^{[m]}
\eea
where $\bp$ is the momentum of the intermediate state. The quantity $E_{\bp}^{[m]}$ is given by
\bea
E_{\bp}^{[m]} = 
\sum_{\nu \neq 0} \!
\frac{\vert \la \nu \vert \sum_\bk \hat{T}_{-\bk+\bp,\bk}^{[n]} \eta_\bk \vert 0 \ra \vert^2}{E_0 - E_\nu}.
\label{Equint}
\eea
Here, $(-\bp)$ is the momentum of the intermediate state $\vert\nu\ra$.
As described in the Section \ref{sec.sqqnt}, the intermediate states $\vert\nu\ra$ that contribute have two triplon quasiparticle excitations and one quintet excitation.
Within the triplet sector, an intermediate state with momentum $(-\bp)$ may be represented as 
\bea
\vert \nu_{2-triplon} \ra = \tau_{\bq-\bp,u'}^\dg \tau_{-\bq,v'}^\dg \vert 0 \ra.
\eea
With this parametrization, the sum over intermediate states $\vert \nu \ra$ may be written as 
\bea
\sum_{\nu \neq 0} \longrightarrow \sum_{\bq}\sum_{u',v'\in\{x,y,z\}}.
\eea
Evaluating the matrix elements using this parametrization of the intermediate state, we find that the energy contribution $E_{\bp}^{[m]}$ is the same from every m-sector, i.e., $E_{\bp}^{[m]} = E_\bp$ for all $m$. The quantity $E_\bp$ is given by
\begin{widetext}
\bea
E_{\bp} = -2\sum_{\bq}\frac{ 
\Big[ 
\sinh^2 (\theta_\bq) \eta_{\bp-\bq}^2 \{ \cosh(2\theta_{\bp-\bq}) + \sinh(2\theta_{\bp-\bq}\} 
+\sinh^2 (\theta_{\bp-\bq}) \eta_{\bq}^2
\{ \cosh(2\theta_\bq) + \sinh(2\theta_{\bq}) \}
\Big]
}{\varepsilon_{q}-\mu+\lambda_{-\bq}+\lambda_{-\bp+\bq}}
\eea
\end{widetext}

\section{Honeycomb bilayer: inclusion of quintets}
\label{app.hcquintet}

In the main text, we defined $\hat{R}_{ttq}(S^2)$ in terms of triplet bilinears  $\hat{A}_{i,i+\delta}^{[n]}$ and $\hat{B}_{i-\delta,i}^{[n]}$. Here, we give expressions for $\hat{A}_{i,i+\delta}^{[n]}$ and $\hat{B}_{i-\delta,i}^{[n]}$ in momentum space. We use the Fourier transform convention 
\bea
t_{i,\alpha\in\{ A,B\},u\in\{x,y,z\}}=\frac{1}{\sqrt{N_\perp /2}}\sum_\bk
t_{\alpha,\bk,u}e^{i\bk.\br_i}.
\eea

(i) The terms in $\hat{A}_{i,i+\delta}^{[n]}$ are of the form $t_{i,A,u}(t_{i+\bdelta,B,v}+ t_{i+\bdelta,B,v}^\dg)$. Using our Fourier transform convention, we may write
\bea
\sum_\bdelta \hat{A}_{i,i+\bdelta}^{[n]} = \frac{M}{N_\perp/2}
\sum_{\bk,\bp} \hat{A}_{-\bk+\bp,\bk}^{[n]}e^{i\bp\cdot \br_i} \gamma_{\bk},
\eea
where $\gamma_\bk = \sum_{\bdelta}e^{i\bk\cdot\bdelta} = 1 + e^{-ik_b} + e^{-ik_a-ik_b}$ and the coefficient $M=\sqrt{\frac{S(S+1)(2S-1)(2S+3)}{30}}$ is the same as that defined for the square lattice case.
%I have taken the factor of $1/sqrt{2}$ into the sq.root.
 The explicit forms of $\hat{A}_{-\bk+\bp,\bk}^{[n]}$ are the same of those of $\hat{T}_{-\bk+\bp,\bk}^{[n]}$ given in Eq.~\ref{Texpns} with the following redefinition: 
\bea
\nn \tilde{t}_{\bk,u} \equiv t_{A,\bk,u} \\
t_{\bk,u} \equiv t_{B,\bk,u}
\eea
 
(ii) The terms in $\hat{B}_{i,i-\bdelta}^{[n]}$ are of the form $t_{i,B,u}( t_{i-\bdelta,A,v} \pm t_{i-\bdelta,A,v}^\dg )$. Using our Fourier transform convention, we write 
\bea
\sum_{\delta} \hat{B}_{i,i-\bdelta}^{[n]} = \frac{M}{N_\perp/2}
\sum_{\bk,\bp} \hat{B}_{-\bk+\bp,\bk}^{[n]} e^{i\bp\cdot \br_i} \gamma_{-\bk}
\eea
Explicit expressions for $\hat{B}_{-\bk+\bp,\bk}^{[n]}$ are the same as those of $\hat{T}_{-\bk+\bp,\bk}^{[n]}$ given in Eq.~\ref{Texpns} but with the following redefinition:
\bea
\nn \tilde{t}_{\bk,u} \equiv t_{B,\bk,u}\\
t_{\bk,u} \equiv t_{A,\bk,u}
\eea
The quintet energy correction on the honeycomb lattice may be rewritten as
\bea
\Delta E^{\rm (q)} = \frac{M^2 \sbar^2}{N_\perp/2} 
\sum_{\bp}\sum_{m}\left[ (A_\bp^{[m]})
+(B_\bp^{[m]}) \right],
\label{Equintfinal}
\eea
where
\bea
\nn (A_\bp^{[m]})=\sum_{\nu\neq 0}\frac{\vert  \la \nu \vert \sum_{\bk} \hat{A}_{-\bk+\bp,\bk}^{[m]} \gamma_{\bk} \vert 0 \ra \vert^2
}{E_0-E_\nu} \\
(B_\bp^{[m]}) = \sum_{\nu\neq 0}\frac{
\vert  \la \nu \vert \sum_{\bk} \hat{B}_{-\bk+\bp,\bk}^{[m]} \gamma_{-\bk} \vert 0 \ra \vert^2
}{E_0-E_\nu}
\label{Eq:ABoverlaps}
\eea
The only intermediate states $\vert \nu \ra$ that contribute to the energy are states with two triplon quasiparticle excitations. 
A generic intermediate state with momentum $(-\bp)$ may be characterized as
\bea
\vert \nu_{2-triplon}\ra = \vartheta_{-\bq,f,u}^\dg \vartheta_{\bq-\bp,g.v}^\dg 
\vert 0 \ra.
\eea
Using this parametrization of a generic state, the sum over intermediate states in Eq.~\ref{Eq:ABoverlaps} becomes
\bea
\sum_{\nu\neq 0 } \longrightarrow \frac{1}{2}  \sum_{\bq}\sum_{f,g\in\{1,2\}}\sum_{u,v\in \{x,y,z\}},
\eea
There is a factor of $1/2$ to account for double counting as $(\bq'= \bp-\bq, f'=g, g'=f)$ corresponds to the same state as $(\bq,f,g)$.
Evaluating the necessary overlaps, we find that the contribution from each m is the same $(A_\bp^{[m]})=(B_\bp^{[m]})=E_\bp$ for $m=-2,\cdots,2$. 
 The quantity $E_\bp$ is given by
\begin{widetext}
\bea
E_\bp = -2\sum_{\bq,f,g} \frac{ \Big[
S_{\bq,f}(-1)^g (S_{-\bp+\bq,g}+C_{\bp-\bq,g})\vert \gamma_{\bp-\bq} \vert
+ S_{\bp-\bq,g}(-1)^f (S_{-\bq,f}+C_{\bq,f})\vert \gamma_{\bq}\vert \Big]^2 }{\varepsilon_{q}-\mu+\lambda_{-\bq,f}+\lambda_{\bq-\bp,g}}
\eea
\end{widetext}
By plugging these expressions into Eq.\ref{Equintfinal}, the correction to ground state energy may be computed.

\end{document}